\documentclass[twocolumn,aps,superscriptaddress,showpacs]{revtex4}
\usepackage{amssymb}
\usepackage{amsmath}
\usepackage{graphicx}
\usepackage[normalem]{ulem}
\usepackage[dvips]{color}

\begin{document}

\title{Imprints of the nuclear symmetry energy on gravitational waves from the axial w-modes of neutron stars}

\author{De-Hua Wen}
\affiliation{Department of Physics, South China University of
Technology,Guangzhou 510641, P.R. China} \affiliation{Department of
Physics and Astronomy, Texas A\&M University-Commerce, Commerce,
Texas 75429-3011, USA}
\author{Bao-An Li\footnote{Corresponding author, Bao-An\_Li$@$Tamu-Commerce.edu}}
\affiliation{Department of Physics and Astronomy, Texas A\&M
University-Commerce, Commerce, Texas 75429-3011, USA}
\author{Plamen G. Krastev}
\affiliation{Department of Physics and Astronomy, Texas A\&M
University-Commerce, Commerce, Texas 75429-3011, USA}
\affiliation{Department of Physics, San Diego State University, 5500
Campanile Drive, San Diego, CA 92182-1233, USA}

\date{\today}

\begin{abstract}
The eigen-frequencies of the axial w-modes of oscillating neutron
stars are studied using the continued fraction method with an
Equation of State (EOS) partially constrained by the recent
terrestrial nuclear laboratory data. It is shown that the density
dependence of the nuclear symmetry energy $E_{sym}(\rho)$  affects
significantly both the frequencies and the damping times of these
modes. Besides confirming the previously found universal behavior
of the mass-scaled eigen-frequencies as functions of the
compactness of neutron stars, we explored several alternative
universal scaling functions. Moreover, the $w_{II}$-mode is found
to exist only for neutron stars having a compactness of $M/R\geq
0.1078$ independent of the EOS used.
\end{abstract}


\maketitle

\section{Introduction}
Presently, among the greatest challenges in modern physics are
determining the equation of state (EOS) of dense neutron-rich matter
and detecting gravitational waves - tiny ripples in space-time
predicted by the theory of general relativity. Physics of neutron
stars provides a unified framework for studying both phenomena: on
the one hand neutron stars are natural sites of super-dense matter,
while on the other hand they are one of the major candidates for
emitting gravitational waves in the bandwidth targeted by the ground
based laser interferometric detectors LIGO \cite{s01}, VIRGO (e.g.
\cite{s02}) and GEO \cite{s03}. Besides the gravitational waves from
elliptically deformed pulsars, an important mechanism of
gravitational radiation from a neutron star is its non-radial
oscillations. The latter may provide potentially a unique probe to
the EOS of super-dense neutron-rich matter because the resulting
gravitational waves carry important information about the neutron
star structure \cite{s1,Nil98,Ben05}.  The non-radial neutron star
oscillations could be triggered by various mechanisms such as
gravitational collapse, a pulsar ``glitch" - a sudden disrupt of the
otherwise very regular pulsar period, gravitational spin-orbit
coupling due to a companion in  a close binary orbit, or a
transition to novel phases of matter in the inner core.

In the framework of general relativity, gravitational radiation
damps out the neutron star oscillations. The frequency of the
non-radial oscillations thus becomes ``quasi-normal" (complex)
with a real part representing the actual frequency of the
oscillation and an imaginary part representing the energy losses.
The eigen-frequencies of the quasi-normal modes can be found by
solving the equations describing the non-radial perturbations of a
static neutron star in general relativity \cite{s2,s3}. These
equations are derived by expanding the perturbed tensors in
spherical harmonics. The perturbation equations split into two
sets: one for the axial mode for which the spherical harmonics
transform under parity (i.e., under the transformation
$\theta\rightarrow\pi-\theta$ and $\phi\rightarrow\pi+\phi$ ) as
$(-1)^{l+1}$; and second for the polar mode for which the
spherical harmonics transform under parity as $(-1)^{l}$,
respectively. With the appropriate boundary conditions at the
neutron star center and surface, the solution of the perturbation
equations yields the complex eigen-frequencies. They can be
classified as follows: (1) f-modes, associated with the global
oscillation of the fluid; (2) g-modes, associated with the fluid
buoyancy; and (3) p-modes, associated with the pressure gradient.
These modes exist in both Newtonian gravitational theory and
general relativity. There is still another type of mode that only
exists in general relativity - the so called w-mode associated
with space-time and for which the motion of the fluid is
negligible \cite{s1,s2,s3}. The w-mode is very important for
astrophysical applications since it is related to the space-time
curvature and exists for all relativistic stars, including black
holes. The standard axial w-mode is categorized as $w_{I}$.
Additionally, there exists an interesting family of axial w-modes
 \cite{s4}, categorized as $w_{II}$. According to the work of
Chandrasekhar \& Ferrari \cite{s2,s3}, the axial w-mode is described
by a unique second-order differential equation. Its simple form
makes it possible to conduct detailed numerical studies of its
spectrum. A major characteristic of the axial w-mode is its high
frequency accompanied by very rapid damping. In fact, the w-mode
frequencies are outside the peak of the detector sensitivities of
the currently operating and planned gravitational wave detectors.
Nevertheless, it is theoretically interesting to investigate the
w-mode of neutron star oscillations as it is an important mechanism
for producing gravitational waves \cite{s2,s3,s4,s10,s11}.
Hopefully, the study will help stimulate discussions on detecting
high-frequency gravitational waves with new generations of detectors
in the future.

The gravitational wave frequency of the axial w-mode depends on the
structure and properties of neutron stars \cite{s4}, which are
determined by the EOS of neutron-rich stellar matter. However, at
present time the EOS of matter under extreme conditions (densities,
pressures and isospin asymmetries) is still rather uncertain and
theoretically controversial. One of the main sources of
uncertainties in the EOS of neutron-rich matter is the poorly known
density dependence of the nuclear symmetry energy, $E_{sym}(\rho)$
\cite{s12}. The EOS of neutron-rich nuclear matter plays an
important role not only in many astrophysical objects/processes
\cite{AWS05} but also in heavy-ion collisions especially those
induced by neutron-rich radioactive beams in terrestrial
laboratories \cite{s17}. While heavy-ion collisions are not expected
to create the same matter and conditions as in neutron stars, the
same elementary nuclear interactions are at work in the two cases
\cite{Xu09}. Thus, it is important to examine ramifications of
conclusions regrading the EOS extracted from one field in the other
one. On one hand, extremely impressive progress has been made in
astrophysical observations relevant for constraining the EOS of
nuclear matter. To our best knowledge, however, mainly because of
the low precision associated with the current measurements of
neutron star radii, a non-controversial conclusion on the EOS and
the density dependence of the nuclear symmetry energy has yet to
come from analyzing astrophysical observations. Nevertheless, it is
very interesting to note that since the pioneering work of Lindblom
\cite{Lin92}, a lot of efforts have been devoted to extracting the
underlying EOS by using the technique of inverting the TOV
(Tolman-Oppenheimer-Volkov) equation using the masses ($M$) and
radii ($R$) simultaneously measured accurately for several neutron
stars, see, e.g., refs. \cite{Lat08,Pra09}, for the latest reports.
While the principle of this technique is well demonstrated and
promising, the lack of simultaneously measured ($M, R$) data has so
far hindered the fruitful applications of this technique. In
particular, while the masses in some binary systems, such as the
double neutron star binaries, are well measured, their radii are
unfortunately not precisely known \cite{Lat08,Pra09}. Similarly,
under the premise that the frequencies and damping rates of several
w-modes of a single neutron star are measured, Tsui and Leung
\cite{TL05} investigated an inversion scheme to determine the mass,
radius and density profile of the neutron star. The underlying EOS
can then also be inferred. While the approach is convincingly
promising, it can only be applied after the gravitational wave
astronomy becomes a reality. On the other hand, heavy-ion reactions
especially those induced by radioactive beams provide an alternative
means to constrain the EOS of neutron-rich nuclear matter, see,
e.g., refs.\cite{LiBA98,LiBA01b,Dan02,Bar05,s17}, for reviews. Over
approximately the last 40 years, the nuclear physics community has
made steady progress in constraining the EOS of symmetric nuclear
matter up to several times the normal nuclear matter density.
However, it was only during roughly the last 10 years the nuclear
physics community has made significant progress in constraining the
density dependence of the symmetry energy of neutron-rich nuclear
matter thanks mainly to the rapid technical advantages in
accelerating radioactive beams \cite{s17}. It is particularly
exciting to see that some significant results in constraining the
density dependence of the symmetry energy have beeb reported very
recently \cite{Xiao09,Tsang09,Cen09,Leh09}. Thus, it is interesting
to investigate how the EOS constrained by the available data from
heavy-ion reactions may help limit the ranges of frequencies and
damping times of the w-mode and the strain of other forms of
gravitational waves \cite{Pla08}. In this work we apply an EOS with
its symmetric part constrained up to about 5 times the normal
nuclear matter density and its symmetry energy constrained in the
subsaturation density region by the available heavy-ion reaction
data. We have two major objectives: (1) to investigate to what
extent the nuclear symmetry energy could affect the eigen-frequency
of axial w-mode; and (2) to examine the relationship between the
axial w-mode eigen-frequency and the neutron star gravitational
field. This study is important because the axial w-mode is believed
to be a pure space-time mode, its eigen-frequency is expected to
have a more direct relationship with the properties of gravitational
field, e.g. with the energy of gravitation. Compared to previous
studies on the w-mode in the literature, the major issues we address
here are scientifically important and studied from a different
direction. In light of the great importance of and the extreme
difficulties encountered in detecting gravitational waves, our
results presented here are scientifically useful.

This paper is organized as follows: after the introduction, in
Section II we recall the formalism and the continued fraction
method for computing the eigen-frequency of the axial w-mode (the
set of perturbation equations and their numerical solution); we
then briefly outline the EOSs applied in our studies in Section
III; in Section IV we present our numerical results; and in
Section V we conclude with a short summary.

\section{Formalism for calculating the axial w-mode frequency}

In what follows we summarize the formalism for calculating the
eigen-frequency of the axial w-mode and the continued fraction
method. Unless noted otherwise, we use geometrical unit $(G=c=1)$.
According to Chandrasekhar \& Ferrari \cite{s2,s3}, the axial
perturbation equations for a static neutron star can be simplified
by introducing a function $z(r)$, constructed from the radial part
of the perturbed axial metric components.  It satisfies the
following differential equation
\begin{equation}\label{RW}
\frac{d^{2}z}{dr_{*}^{2}}+[\omega^{2}-V(r)]z=0,
\end{equation}
where $\omega (=\omega_{0}+i\omega_{i})$  is the complex
eigen-frequency of the axial w-mode. Inside the star, the tortoise
coordinate $r_{*}$ and the potential function $V$ are defined by
  \begin{equation}\label{rstar}
 r_{*}=\int_{0}^{r}e^{\lambda-\nu}dr \  (or \
 \frac{d}{dr_{*}}=e^{\lambda-\nu}\frac{d}{dr})
 \end{equation}
and
  \begin{equation}\label{VP}
V=\frac{e^{2\nu}}{r^{3}}[l(l+1)r+4\pi r^{3}(\rho-p)-6m]
 \end{equation}
with $ l$ the spherical harmonics index (used in describing the
perturbed metric, here only the case $l=2$ is considered), $\rho$
and $p$  the density and pressure,  and $m$ the mass inside radius
$r$, respectively. The functions $e^{\nu}$ and $e^{\lambda}$ are
given by the line element for a static neutron star as
 \begin{equation}
-ds^{2}=-e^{2\nu}dt^{2}+e^{2\lambda}dr^{2}+r^{2}(d\theta^{2}+\textrm{sin}^{2}\theta
d\phi^{2}).
 \end{equation}
Outside the neutron star, Eqs.\ \ref{rstar} and \ref{VP} reduce to
 \begin{equation}
 r_{*}=2M \textrm{ln}(r-2M) \  (or \
 \frac{d}{dr_{*}}=\frac{r-2M}{r}\frac{d}{dr})
 \end{equation}
and
  \begin{equation}
V=\frac{r-2M}{r^{4}}[l(l+1)r-6M].
 \end{equation}
 where $M$ is the total gravitational mass of neutron star.
The solutions to this problem are subject to a set of boundary
conditions (BC) constructed by Chandrasekhar \& Ferrari \cite{s2}
- regular BC at the neutron star center, continuous BC at the
surface and behaving as a purely outgoing wave at infinity.

 As discussed in references \cite{s1,s4,s31}, the w-modes have larger imaginary parts and this introduces
 considerable difficulties to dealing with the boundary conditions at infinity. For a larger radius,
 the term representing the incoming wave, $z^{in}\sim e^{-r_{*}\omega_{i}}$, vanishes and therefore the problem no longer satisfies the necessary BC.
 A practical way to overcome these difficulties is to utilize the \textit{continued fraction method} \cite{s4,s31},
 which is applicable to both the axial and polar w-modes. This method provides both the $w_{I}$- and $w_{II}$-modes.

Here we give a brief introduction to the continued fraction method
following closely Benhar et al. \cite{s4}.   It is convenient to
use the dimensionless geometrical units, i.e. $2M=c=G=1$. First,
one represents the solution outside the neutron star as

 \begin{equation}
z(r)=(r-1)^{-i\omega}e^{-i\omega
r}\sum_{n=0}^{\infty}a_{n}y^{n}\equiv\chi(r)\sum_{n=0}^{\infty}a_{n}y^{n},
 \end{equation}
where $y=1-a/r, R<a<2R$ ($R$ is the stellar radius) and the
coefficients $a_{n}$ satisfy the following four-term recurrence
relation
\begin{equation} \label{recu}
 \alpha_{n}a_{n+1}+\beta_{n}a_{n}+\gamma_{n}a_{n-1}+\delta_{n}a_{n-2}=0;\
 (n\geq 2)
 \end{equation}
with
\begin{equation}
 \alpha_{n}=(1-\frac{1}{a})n(n+1),
 \end{equation}
\begin{equation}
 \beta_{n}=-2(i\omega a+n-\frac{3n}{2a})n,
 \end{equation}
\begin{equation}
 \gamma_{n}=(1-\frac{3}{a})n(n-1)+\frac{3}{a}-l(l+1),
 \end{equation}
\begin{equation}
 \delta_{n}=\frac{1}{a}(n-3)(n+1).
 \end{equation}
The coefficients $a_{0}$ and $a_{1}$ are determined by the
continuity of $z(r)$ and $z(r)_{,r}$ at $r=a$, that is
\begin{equation} \label{a0}
  a_{0}=\frac{z(a)}{\chi(a)},
 \end{equation}
\begin{equation} \label{a1}
  a_{1}=\frac{a}{\chi(a)}[z_{,r}(a)+\frac{i\omega a}{a-1}z(a)].
 \end{equation}
Since  Eq.\ \ref{recu}  is a four-term recurrence relation, to
determine $a_{n}$ uniquely one must know three initial terms, i.e.
 Eqs.\ \ref{a0} and \ref{a1} are not sufficient alone. Leaver \cite{s32} has
shown that  Eq.\ \ref{recu} can be  further reduced to a three-term
recurrence relation as
\begin{equation}
 \hat{\alpha}_{n}a_{n+1}+\hat{\beta}_{n}a_{n}+\hat{\gamma}_{n}a_{n-1}=0,\
 (n\geq 2),
 \end{equation}
for $n=0$
\begin{equation}
 \hat{\alpha}_{0}=-1, \ \hat{\beta}_{0}=\frac{a_{1}}{a_{0}},
 \end{equation}
for $n=1$
\begin{equation}
 \hat{\alpha}_{1}=\alpha_{1}, \ \hat{\beta}_{1}=\beta_{1},\
 \hat{\gamma}_{1}=\gamma_{1},
 \end{equation}
for $n\geq 2$
\begin{equation}
 \hat{\alpha}_{n}=\alpha_{n}, \ \hat{\beta}_{n}=\beta_{n}-\frac{\hat{\alpha}_{n-1}\delta_{n}}{\hat{\gamma}_{n-1}},\
 \hat{\gamma}_{n}=\gamma_{n}-\frac{\hat{\beta}_{n-1}\delta_{n}}{\hat{\gamma}_{n-1}}.
 \end{equation}
According to Leaver \cite{s32} and Wall \cite{s33}, the function
$z(r)$ describes a purely outgoing wave at infinity only if
\begin{widetext}
\begin{equation} \label{frac}
 f_{n}(\omega)= \hat{\beta}_{n}-\frac{\hat{\alpha}_{n-1}\hat{\gamma}_{n}}{\hat{\beta}_{n-1}-\frac{\hat{\alpha}_{n-2}\hat{\gamma}_{n-1}}{\hat{\beta}_{n-2}-...\frac{\hat{\alpha}_{0}\hat{\gamma}_{1}}{\hat{\beta}_{0}}}}
 -\frac{\hat{\alpha}_{n}\hat{\gamma}_{n+1}}{\hat{\beta}_{n+1}-\frac{\hat{\alpha}_{n+1}\hat{\gamma}_{n+2}}{\hat{\beta}_{n+2}-\frac{\hat{\alpha}_{n+2}\hat{\gamma}_{n+3}}{\hat{\beta}_{n+3}-...}}}=0 \
 (n=0,1,2,...).
 \end{equation}
\end{widetext}
For $n=0$, the above expression becomes
\begin{equation} \label{frac0}
 f_{0}(\omega)= \hat{\beta}_{0}-\frac{\hat{\alpha}_{0}\hat{\gamma}_{1}}{\hat{\beta}_{1}-
 \frac{\hat{\alpha}_{1}\hat{\gamma}_{2}}{\hat{\beta}_{2}-\frac{\hat{\alpha}_{2}\hat{\gamma}_{3}}{\hat{\beta}_{3}-...}}}=0.
 \end{equation}
Eq.\ \ref{frac} is the key expression for computing the
eigen-frequencies of the axial w-modes. Obviously
$f_{0}(\omega)=0$ is the simplest case, which is mainly employed
in our calculation.  In order to rule out the pseudo roots, the
cases $n=1$ and 2 are also examined.

 The numerical procedure goes as follows: (1) choose a suitable range for $\omega_{0}$  and $\omega_{i}$;
 (2) integrate Eq.\ \ref{RW} starting at the center to $r=a$;
 (3) compute  $f_{0}(\omega)$ through Eq.\ \ref{frac0} and plot the curve $\omega_{0}$  against $\omega_{i}$
  (where the real part and the imaginary part of function  $f_{0}(\omega)$  are
  zero); (4)   determine the crossing point of the curves - the crossing point specifies
   the eigen-frequency of the axial w-mode. The same procedure is also repeated for the cases with $n=1$ and $n=2$.

\section{The equation of state of neutron-rich nuclear matter partially constrained by recent terrestrial nuclear laboratory data}

The necessary input in solving the Eq.\ \ref{RW} is the EOS. While
at high densities reached in the core of neutron stars, other
particles and new phases of matter may exist, in this work we
consider the simplest model of neutron stars consisting of neutrons,
protons and electrons ($npe$) in beta-equilibrium. For isospin
asymmetric nuclear matter, various theoretical studies have shown
that the energy per nucleon can be well approximated by
\begin{equation}
E(\rho,\delta)=E(\rho,\delta=0)+E_{sym}(\rho)\delta^{2}+O(\delta^{4})
\end{equation}
in terms of the baryon density $\rho=\rho_{n}+\rho_{p}$, the
isospin asymmetry
$\delta=(\rho_{n}-\rho_{p})/(\rho_{n}+\rho_{p})$, the energy per
nucleon in symmetric nuclear matter $E(\rho,\delta=0)$, and the
bulk nuclear symmetry energy $E_{sym}(\rho)$
\cite{s12,s24,s25,s26}. The corresponding pressure of the $npe$
matter is given by
\begin{eqnarray}\label{pre}
P(\rho,\delta)&=&\rho^2\left(\frac{\partial E}{\partial \rho}
\right)_{\delta}+\frac{1}{4}\rho_e\mu_e\nonumber\\
&=&\rho^2\left[E'(\rho,\delta=0)+E'_{\rm sym}(\rho)\delta^2\right]\nonumber\\
&+&\frac{1}{2}\delta(1-\delta)\rho E_{\rm sym}(\rho),
\end{eqnarray}
where $\rho_e=\frac{1}{2}(1-\delta)\rho$ and
$\mu_e=\mu_n-\mu_p=4\delta E_{\rm sym}(\rho)$ are, respectively,
the density and chemical potential of electrons. It is obvious
that the pressure is dominated by the symmetry energy term in the
$npe$ matter near the saturation density where the
$E'(\rho,\delta=0)$ term vanishes. The value of the isospin
asymmetry $\delta$ at $\beta$ equilibrium is determined by the
chemical equilibrium and charge neutrality conditions, i.e.,
$\delta=1-2x_p$ with

\begin{equation}
x_p\approx 0.048 \left[E_{\rm sym}(\rho)/E_{\rm
sym}(\rho_0)\right]^3 (\rho/\rho_0)(1-2x_p)^3.
\end{equation}

It is seen that the proton fraction $x_p$ is uniquely determined
by the density dependence of the nuclear symmetry energy
$E_{sym}(\rho)$. It is well-known that while the maximum mass of
neutron stars is determined mostly by the stiffness of the
symmetric nuclear EOS $E(\rho,\delta=0)$, the radii of neutron
stars are mostly determined by the slope of the $E_{sym}(\rho)$
\cite{s12}. One of the most uncertain part of the EOS of
neutron-rich nuclear matter is the density dependence of the
$E_{sym}(\rho)$ especially at supra-saturation densities
\cite{s17}. In the previous studies of the eigen-frequencies of
the w-mode using various EOSs \cite{Nil98,s4,s35,s36}, the focus
was on exploring effects of the stiffness of the EOS and
understanding the scaling behavior of the eigen-frequencies. It is
still not clear what are the effects of the $E_{sym}(\rho)$. In
this study we use EOSs with the same incompressibility $K$ at
normal density but different $E_{sym}(\rho)$. Our study is thus
complementary to the existing studies \cite{Nil98,s4,s35,s36}.
Moreover, the EOS of neutron-rich nuclear matter also plays an
important role in heavy-ion collisions especially those induced by
neutron-rich radioactive beams in terrestrial laboratories.
Significant progress has been made recently in constraining the
EOS of neutron-rich nuclear matter using heavy-ion experiments,
see, e.g., refs.~\cite{Dan02,Bar05,s17}. In particular,
experimental data on collective flow and kaon production in
relativistic heavy-ion collisions have put a strong constraint on
the EOS of symmetric nuclear matter at densities up to about 5
times the normal nuclear matter density \cite{Dan02}. On the other
hand, the $E_{sym}(\rho)$ in the sub-saturation density region has
also been constrained recently by analyzing data on isospin
diffusion \cite{s18,s19,s20,s21} and isoscaling \cite{s22,s23} in
heavy-ion reactions at intermediate energies. In this work, we use
the MDI (Momentum Dependent Interaction) EOSs obtained from
Hartree-Fock calculations using a modified Gogny interaction
\cite{s018}. The parameters of the interactions are adjusted such
that the resulting EOSs satisfy all of the above constraints from
heavy-ion reactions.

For the modified Gogny MDI interaction, the baryon potential energy density can be
expressed as~\citep{s018}
\begin{widetext}
\begin{equation}
V(\rho,\delta ) =\frac{A_{u}(x)\rho _{n}\rho _{p}}{\rho _{0}}
+\frac{A_{l}(x)}{2\rho _{0}}(\rho _{n}^{2}+\rho
_{p}^{2})+\frac{B}{\sigma +1}\frac{\rho ^{\sigma +1}}{\rho
_{0}^{\sigma }} (1-x\delta ^{2})+\frac{1}{\rho _{0}}\sum_{\tau
,\tau ^{\prime}}C_{\tau ,\tau ^{\prime }} \int \int d^{3}pd^{3}p^{\prime }\frac{f_{\tau }(\vec{r},\vec{p}%
)f_{\tau ^{\prime }}(\vec{r},\vec{p}^{\prime
})}{1+(\vec{p}-\vec{p}^{\prime })^{2}/\Lambda ^{2}}.
 \label{MDIVB}
\end{equation}
\end{widetext}

\begin{figure}
\includegraphics{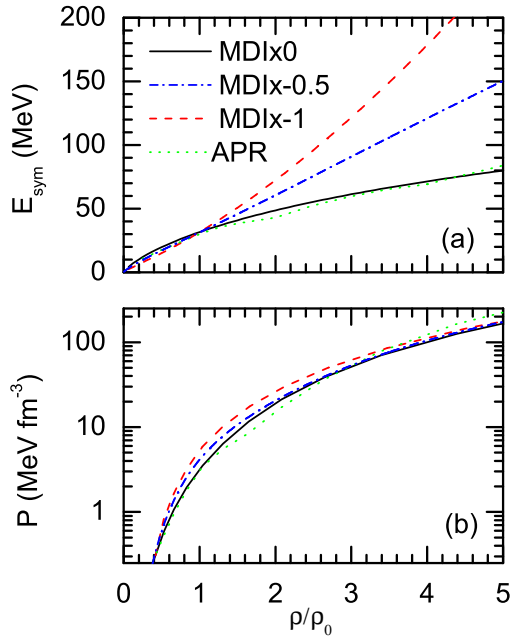}
\caption{\label{fig1} (Color online) The symmetry energy (a) and
pressure (b) versus the baryon number density (baryon number
density $\rho$ is in units of saturation nuclear density
$\rho_{0}$).}
\end{figure}

\begin{figure}
\includegraphics{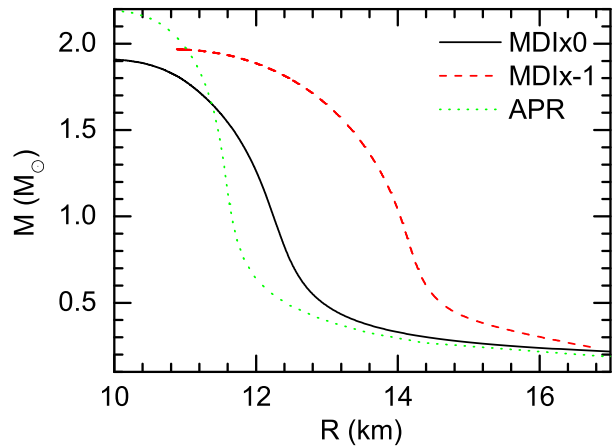}
\caption{\label{fig2}  (Color online) Mass-radius relation for
static neutron stars.}
\end{figure}

In the above equation, the isospin $\tau =1/2$ ($-1/2$) is for
neutrons (protons). The coefficients $A_{u}(x)$ and $A_{l}(x)$
are~\cite{s20}
\begin{equation}\label{alau}
A_{u}(x)=-95.98-x\frac{2B}{\sigma
+1},~~~~A_{l}(x)=-120.57+x\frac{2B}{\sigma +1}.
\end{equation}
The values of the parameters are $\sigma=4/3$, $B=106.35$ MeV,
$C_{\tau ,\tau }=-11.70$ MeV, $C_{\tau ,-\tau }=-103.40$ MeV and
$\Lambda=p_f^{0}$ which is the Fermi momentum of nuclear matter at
$\rho_0$. The parameter $x$ was introduced to mimic various
$E_{sym}(\rho)$ predicted by different microscopic many-body
theories. By adjusting the $x$ parameter, the $E_{sym}(\rho)$ is
varied without changing any property of symmetric nuclear matter
and the symmetry energy at saturation density as the $x$-dependent
$A_{u}(x)$ and $A_{l}(x)$ are automatically adjusted accordingly.
We note especially that the symmetry energy at normal density
$E_{sym}(\rho_0)$ is fixed at 31.6 MeV and the incompressibility
$K$ of symmetric nuclear matter at normal density is fixed at 211
MeV consistent with known experimental constraints \cite{s17}. We
stress that both the $E_{sym}(\rho_0)$ and $K$ are independent of
the $x$ parameter. It was shown that \cite{s21,s27} only values of
$x$ in the range between $-1$ (MDIx-1) and $0$ (MDIx0) are
consistent with the isospin-diffusion and isoscaling data as well
as the available measurements of the neutron-skin thickness of
$^{208}$Pb. These data only constrain the $E_{sym}(\rho)$ at
sub-saturation densities. At supra-saturation densities, some
experimental indications on the trend of the $E_{sym}(\rho)$ has
just start to appear\cite{Xiao09}. For this exploratory study we
assume that the two limiting functions of $E_{sym}(\rho)$ with
$x=0$ and $x=-1$ experimentally constrained only at sub-saturation
densities can be extrapolated to supra-saturation densities
according to the MDI predictions. Hopefully, future theoretical
and experimental studies will allow us to better constrain also
the high-density symmetry energy and thus relax the above
assumption. Moreover, as we indicated earlier, extrapolating
hadronic EOS to high densities without considering the possible
hadron-QGP (Quark-Gluon-Plasma) phase transition is always a
problem for neutron star models. Given our limited knowledge about
the hadron-QGP phase transition and other properties of
super-dense matter, for our purpose of studying the imprints of
nuclear symmetry energy on the w-mode, we stick to the simplest
model of neutron stars made of the $npe$ matter.

For comparisons, we also employ the widely used EOS by Akmal,
Pandharipande and Ravenhall (APR) \cite{s28}. Shown in   Fig.\
1(a) is the nuclear symmetry energy considered in this work. It is
interesting to note that the APR prediction on the $E_{sym}(\rho)$
at sub-saturation densities lies right between the MDIx0 and
MDIx-1 predictions. At supra-saturation densities it follows
closely the MDI $E_{sym}(\rho)$ with $x=0$. At nuclear densities
below approximately $0.07 ~\textrm{fm}^{-3}$, we supplement the
EOSs by those of Refs. \cite{s29,s30} which are more suitable for
the neutron star crust. The saturation properties of symmetric
nuclear matter for the EOSs used here are summarized in Table I.
We emphasize that the APR predicts a significantly stiffer
incompressibility of 266 MeV compared to the MDI EOS which has an
incompressibility of 211 MeV independent of the symmetry energy
parameter $x$.

\begin{table}
\caption{\label{tab:table1} Saturation properties of the symmetric
nuclear matters EOS. The first column identifies the EOS. The
remaining columns presenting the following quantities at nuclear
saturation density: saturation baryon density; energy per particle;
incompressibility; nucleon effective mass; symmetry energy. }
\begin{ruledtabular}
\begin{tabular}{cccccc}
EOS & $\rho_{0}(fm^{-3})$ & $E_{s}(MeV)$ & $K(MeV)$ & $m(MeV/c^{2})$ & $E_{sym}(MeV)$\\
\hline
MDI & 0.160 & -16.08 & 211.00 & 629.08 & 31.62\\
APR & 0.160 & -16.00 & 266.00 & 657.25 & 32.60\\
 \end{tabular}
\end{ruledtabular}
\end{table}

Shown in Fig.\ 1(b) is the pressure of the $npe$ matter as a
function of the baryon density. As one expects, the stiffer
symmetry energy with $x=-1$ leads to a higher pressure compared to
$x=0$. As a result, the radii of neutron stars are consistently
larger with $x=-1$ as shown in Fig.\ 2. We notice that the MDIx0
and MDIx-1 EOSs lead to about the same maximum mass for neutron
stars. This is because they have the same incompressibility $K$ of
211 MeV for symmetric nuclear matter at normal density. The APR
EOS leads to a significantly larger maximum mass because of its
stiffer incompressibility but similar radii as the MDIx0
prediction because of the similar symmetry energy density
functionals near the saturation density. We notice here that the
range of radii between the MDIx0 and MDIx-1 predictions is much
smaller than that spanned by the various EOSs considered in ref.
\cite{s12}. In the latter the EOSs are different not only in their
symmetry energy functionals but also the incompressibilities for
symmetric nuclear matter.

\section{Results and discussion}
In this section we present our numerical results for the first axial
w-mode ($w_{I}$-mode), the  $w_{II}$-mode, the second w-mode
($w_{I2}$-mode), and some of the third axial w-modes (
$w_{I3}$-modes). The calculation is performed applying the continued
fraction method together with the EOSs we discussed in the previous
Section.

We first examine effects of the symmetry energy on the frequencies
and damping times of the w-mode. Shown in Figs. 3 and 4 are the
frequency and damping time of the $w_{I}$-mode (a) and $w_{II}$-mode
(b) respectively, as functions of the neutron star mass. These
figures establish the relationship between the expected frequencies
of the axial w-modes, for a given EOS, and the stellar mass. It is
interesting to notice, in Fig.\ 3, that there is a clear difference
between the frequencies calculated with the MDIx0 EOS and those with
the MDIx-1 EOS. Since the major difference between these two cases
is the density dependence of the nuclear symmetry energy, it is
obvious that the symmetry energy has a clear imprint on the
frequencies. Therefore, the symmetry energy constrained by nuclear
reactions in terrestrial laboratories can help determine the
expected frequency of the axial w-mode and its damping time.

\begin{figure}
\includegraphics{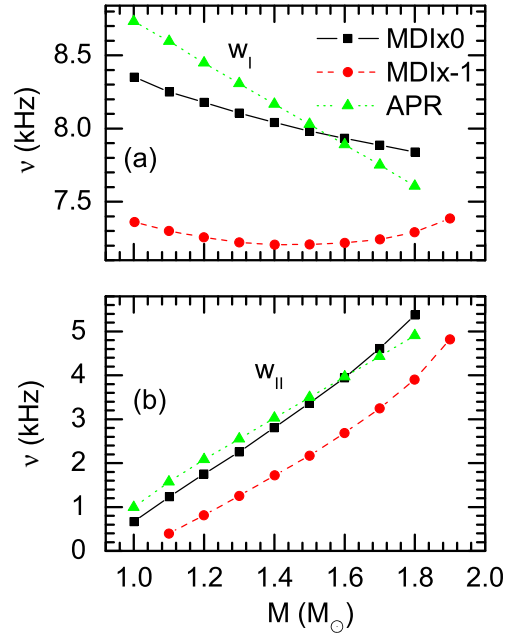}
\caption{\label{fig3} (Color online) Frequency of $w_{I}$-mode (a)
and $w_{II}$-mode  (b) as a function of the neutron star mass
$M$.}
\end{figure}

\begin{figure}
\includegraphics{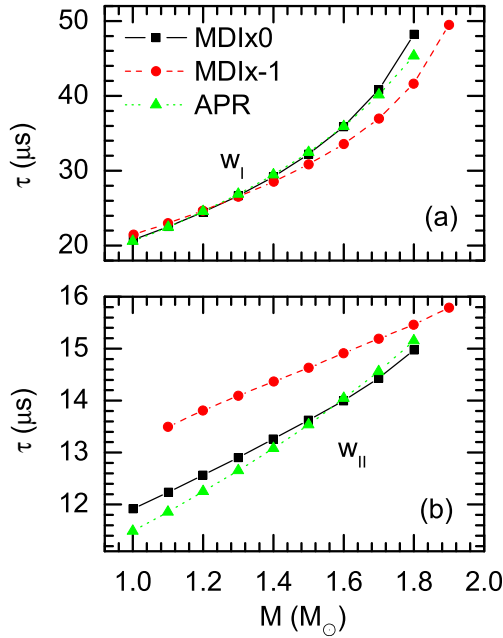}
\caption{\label{fig4} (Color online) Damping time of $w_{I}$-mode
(a)  and $w_{II}$-mode (b) as a function of the neutron star mass
$M$.}
\end{figure}

\begin{figure}
\includegraphics{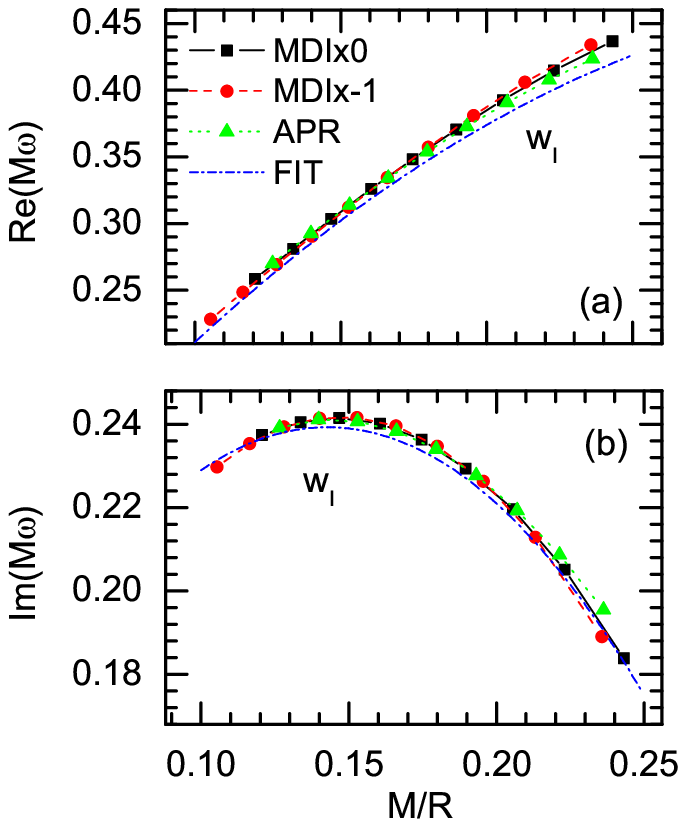}
\caption{\label{fig5}(Color online) Eigen-frequency, $\omega$, of
the $w_{I}$-mode scaled by the stellar mass $M$ as a function of
compactness $M/R$. The fit is performed by employing the
parameters of Tsui et al. \cite{s35}.}
\end{figure}

\begin{figure}
\includegraphics{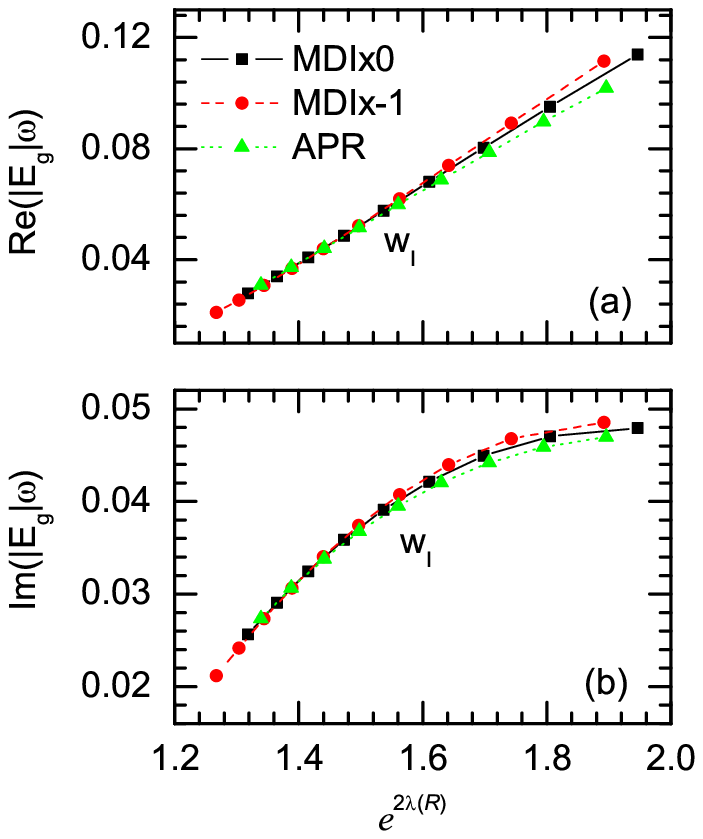}
\caption{\label{fig6}(Color online) Eigen-frequency, $\omega$, of
the $w_{I}$-mode scaled by the gravitational energy  $|E_{g}|$
versus the metric function $e^{2\lambda}$ at the stellar surface.}
\end{figure}

\begin{figure}
\includegraphics{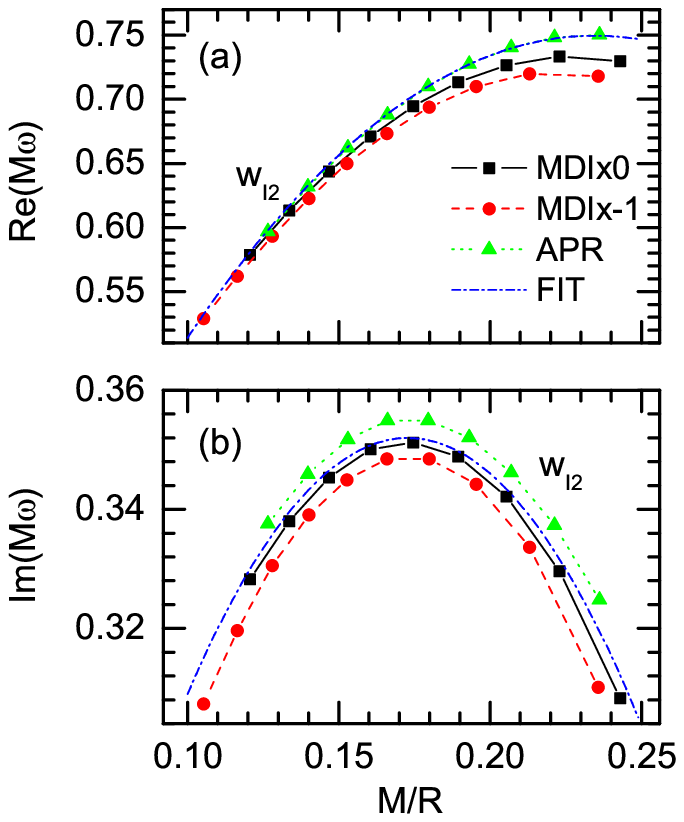}
\caption{\label{fig7}(Color online) Eigen-frequency, $\omega$, of
the $w_{I2}$-mode scaled by the stellar mass $M$ as a function of
compactness $M/R$. The fit is performed by employing the
parameters of Tsui et al. \cite{s35}.}
\end{figure}

\begin{figure}
\includegraphics{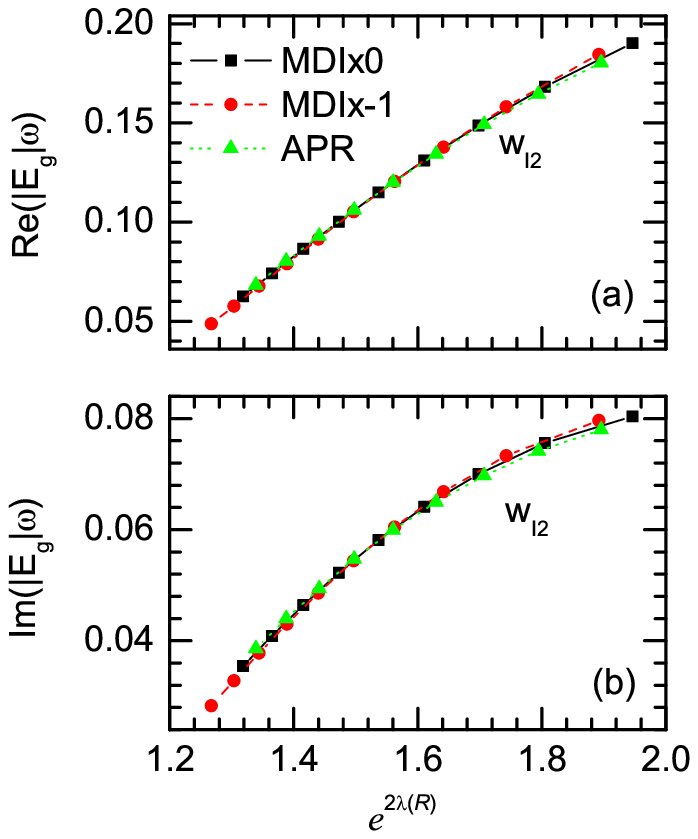}
\caption{\label{fig8}(Color online) Eigen-frequency, $\omega$, of
the $w_{I2}$-mode scaled by the gravitational energy  $|E_{g}|$
versus the metric function $e^{2\lambda}$ at the stellar surface.}
\end{figure}

\begin{figure}
\includegraphics{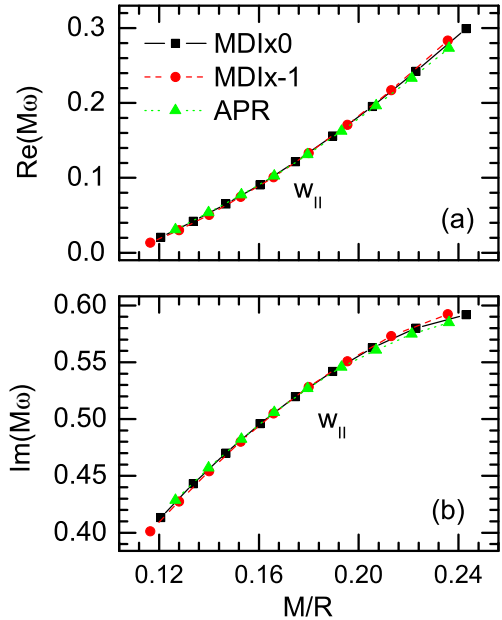}
\caption{\label{fig9}(Color online) Eigen-frequency, $\omega$, of
the $w_{II}$-mode scaled by the stellar mass $M$ as a function of
compactness $M/R$. }
\end{figure}

\begin{figure}
\includegraphics{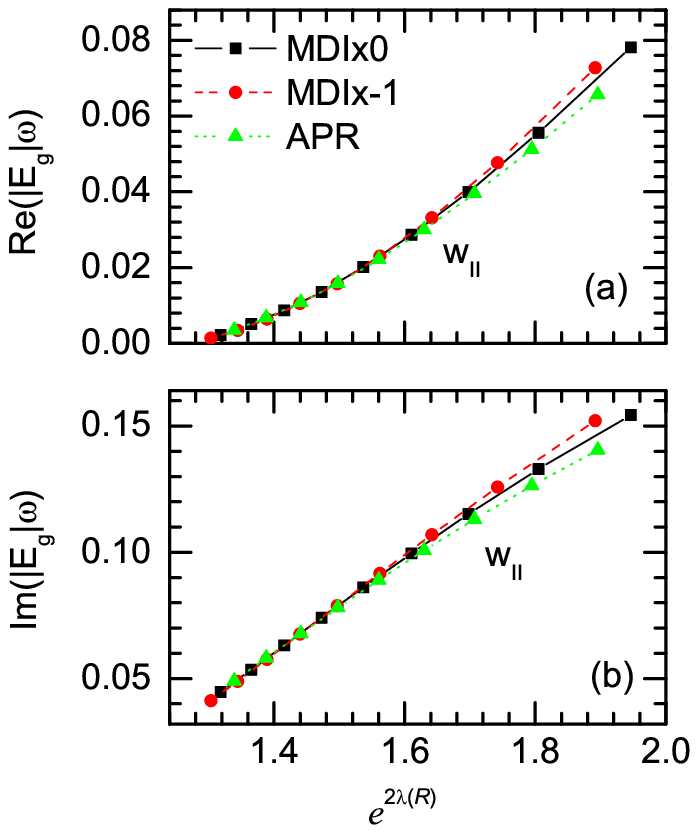}
\caption{\label{fig10} (Color online) Eigen-frequency, $\omega$,
of the $w_{II}$-mode scaled by the gravitational energy $|E_{g}|$
versus the metric function $e^{2\lambda}$ at the stellar surface.}
\end{figure}

In Figures 5, 7 and 9 we show the real (a) and imaginary (b) parts
of the eigen-frequency   of $w_{I}$-, $w_{I2}$- and $w_{II}$-modes
scaled by the mass $M$ as a function of the neutron star
compactness $M/R$, respectively.  These results suggest that the
scaled eigen-frequency exhibits a universal behavior independent
of the EOS used as a function of the compactness parameter.
Similar universal behaviors were first found for the polar w-mode
by Andersson et al.\cite{Nil98} and later for the axial w-mode by
Benhar et al. \cite{s34} and Tsui et al. \cite{s35,s36}. As it was
pointed out earlier in the above references, provided that the
masses and radii of neutron stars are known, the universal scaling
behaviors allows an accurate determination of the w-mode frequency
and damping time of gravitational waves. This is very important
for guiding the gravitational wave search. On the other hand, if
both the frequency and damping time for a given neutron star are
known this could provide information on the neutron star mass and
radius. In Figs.\ 5 and 7 we also display the curves (Fit) best
fitting the results obtained by Tsui et al. using eight different
EOSs \cite{s35}. It is seen that our numerical results are in good
agreement with theirs.
\begin{figure}
\includegraphics{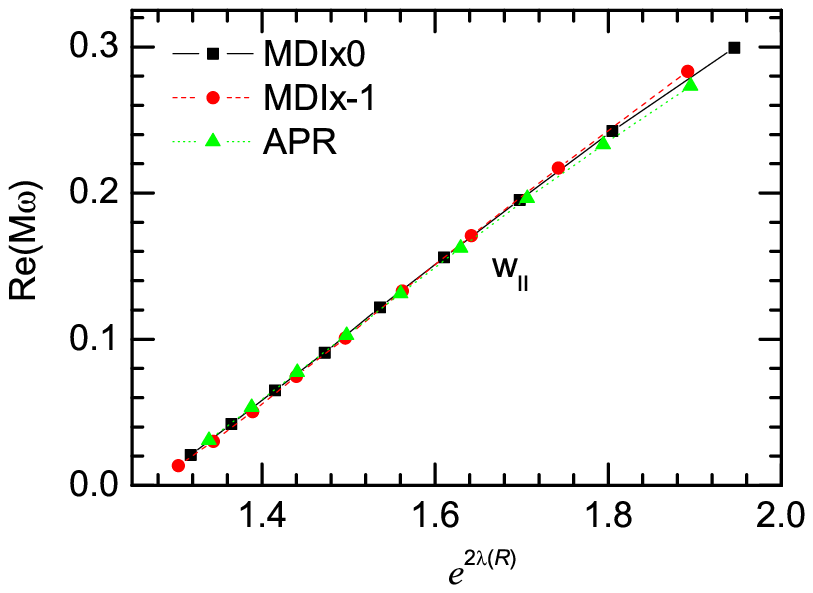}
\caption{\label{fig11}(Color online) Real part of the scaled
frequency $M\omega$ of $w_{II}$-mode versus the metric function
$e^{2\lambda}$ at the stellar surface. }
\end{figure}

\begin{figure}
\includegraphics{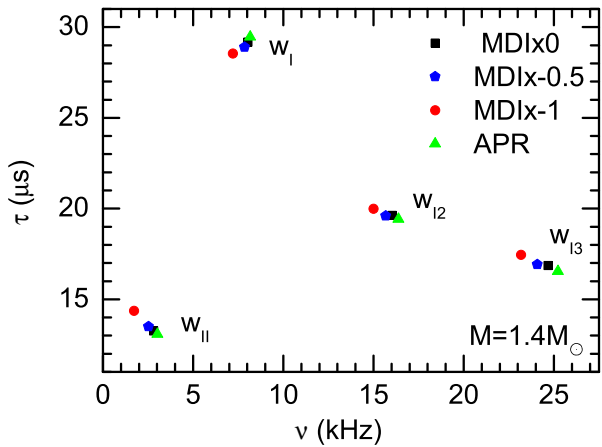}
\caption{\label{fig12} (Color online) Frequency and damping time
of the first, second and third axial w-mode ($w_{I}$, $w_{I2}$ and
$w_{I3}$) and the axial $w_{II}$-mode for neutron star models of
mass $M=1.4M_{\odot}$. }
\end{figure}

Realizing that the axial w-mode is a space-time mode \cite{s4}, it
is interesting to study whether the w-mode eigen-frequencies may be
scaled by the total gravitational energy and how the scaled
frequencies behave as functions of the metric function
$e^{2\lambda(R)}=1/(1-2M/R)$ (at the stellar surface). This approach
relates again the eigen-frequency to the compactness parameter $M/R$
of neutron stars, but additionally it has the advantage of taking
into account more completely the space-time properties. The
gravitational energy is calculated from \cite{s37}
 \begin{equation}
 E_{g}=\int_{0}^{R}4\pi r^{2}\{1-[1-\frac{2m(r)}{r}]^{-1/2}\}\rho
 dr.
 \end{equation}
Shown in Figs. 6, 8 and 10 are the real (a) and imaginary (b)
parts of the eigen-frequencies of the $w_{I}$-, $w_{I2}$- and
$w_{II}$-modes scaled by the absolute value of the gravitational
energy $|E_{g}|$ as functions of the metric function
$e^{2\lambda(R)}$ at the surface. Compared to the corresponding
frequencies scaled by the neutron star mass as functions of $M/R$,
while the universal behavior of the $w_{I}$- and $w_{II}$-mode
frequencies remain about the same, the $w_{I2}$-mode frequencies
scaled by the total gravitational energy as functions of
$e^{2\lambda(R)}$ become much more universal.

Besides the above two kinds of universal scalings, it is possible
to obtain alternative scalings. One particularly useful scaling we
observed is that the real frequency $\omega$ of the $w_{II}$-mode
scaled by the mass $M$ varies linearly and independent of the EOS
as a function of the metric function $e^{2\lambda (R)}$, see Fig.\
11. From the lower panel of Fig.\ 3, one can see that there exists
a minimum mass or compactness, below which the $w_{II}$-mode
frequency will vanish, namely, there is no $w_{II}$-mode below the
limit. Unfortunately, it is not easy to obtain this limit directly
based on Fig.\ 3. However, the linear characteristic shown in
Fig.\ 11 allows one to easily and accurately determine the minimum
compactness for the existence of the $w_{II}$-mode to be
$M/R\approx 0.1078$.

To be more quantitative and make it easier to compare with future
studies, we summarize in Table II several neutron star properties,
the corresponding eigen-frequencies and damping times for the EOSs
considered in this work.

\begin{table*}
\caption{\label{tab:table3}
 Neutron star properties (central density, radius, mass, compactness and gravitational energy),
  and eigen-frequencies and damping times of the $w_{I}$- and $w_{II}$- modes for the EOSs considered in this paper.}
\begin{ruledtabular}
\begin{tabular}{c|ccccccccc}

  EOS & $\rho_{c}(10^{18}kg\cdot m^{-3})$ & $R(km)$ & $M(M_{\odot})$ & $M/R$ & $|E_{g}|(M_{\odot})$ & $\nu_{w_{I}}(kHz)$ & $\tau_{w_{I}}(\mu s)$ & $\nu_{w_{II}}(kHz)$ & $\tau_{w_{II}}(\mu s)$ \\ \hline
  & 0.7265 & 12.25 & 1.00 & 0.121 & 0.108 & 8.35 & 20.8 & 0.67 & 11.9 \\
  & 0.7970 & 12.16 & 1.10 & 0.134 & 0.133 & 8.25 & 22.6 & 1.23 & 12.2 \\
  & 0.8728 & 12.07 & 1.20 & 0.147 & 0.161 & 8.18 & 24.5 & 1.75 & 12.6 \\
  & 0.9586 & 11.96 & 1.30 & 0.160 & 0.194 & 8.11 & 26.7 & 2.26 & 12.9 \\
  MDIx0& 1.0595 & 11.83 & 1.40 & 0.175 & 0.232 & 8.04 & 29.2 & 2.81 & 13.3 \\
  & 1.1780 & 11.68& 1.50 & 0.190 & 0.276 & 7.98 & 32.2 & 3.36 & 13.6 \\
  & 1.3196 & 11.49 & 1.60 & 0.205 & 0.327 & 7.93 & 35.9 & 3.95 & 14.0 \\
  & 1.4969 & 11.25 & 1.70 & 0.223 & 0.390 & 7.89 & 40.8 & 4.61 & 14.4 \\
  & 1.7760 & 10.93 & 1.80 & 0.243 & 0.469 & 7.84 & 48.2 & 5.36 & 15.0  \\  \hline
  & 0.5535 & 13.94 & 1.10 & 0.116 & 0.113 & 7.30 & 23.0 & 0.39 & 13.5  \\
  & 0.6140 & 13.83 & 1.20 & 0.128 & 0.137 & 7.26 & 24.7 & 0.81 & 13.8  \\
  & 0.6835 & 13.70 & 1.30 & 0.140 & 0.165 & 7.22 & 26.5 & 1.25 & 14.1  \\
  & 0.7630 & 13.54 & 1.40 & 0.153 & 0.197 & 7.21 & 28.5 & 1.72 & 14.4  \\
 MDIx-1 & 0.8570 & 13.35 & 1.50 & 0.166 & 0.234 & 7.21 & 30.8 & 2.17 & 14.6  \\
  & 0.9700 & 13.12 & 1.60 & 0.180 & 0.278 & 7.22 & 33.6 & 2.69 & 14.9  \\
  & 1.1181 & 12.84 & 1.70 & 0.195 & 0.330 & 7.24 & 37.0 & 3.25 & 15.2  \\
  & 1.3220 & 12.47 & 1.80 & 0.213 & 0.395 & 7.29 & 41.6 & 3.90 & 15.5  \\
  & 1.6760 & 11.90 & 1.90 & 0.236 & 0.488 & 7.39 & 49.5 & 4.82 & 15.8  \\ \hline
  & 0.7880 & 11.66 & 1.00 & 0.127 & 0.114 & 8.73 & 20.6 & 0.99 & 11.5  \\
  & 0.8383 & 11.62 & 1.10 & 0.140 & 0.140 & 8.60 & 22.5 & 1.57 & 11.9  \\
  & 0.8895 & 11.58 & 1.20 & 0.153 & 0.169 & 8.45 & 24.6 & 2.08 & 12.2  \\
  & 0.9422 & 11.55 & 1.30 & 0.166 & 0.201 & 8.31 & 26.9 & 2.55 & 12.7  \\
  APR& 1.0019 & 11.51 & 1.40 & 0.180 & 0.237 & 8.17 & 29.5 & 3.03 & 13.1  \\
  & 1.0665 & 11.46 & 1.50 & 0.193 & 0.277 & 8.03 & 32.4 & 3.50 & 13.5  \\
  & 1.1370 & 11.41 & 1.60 & 0.207 & 0.322 & 7.89 & 35.9 & 3.97 & 14.0  \\
  & 1.2155 & 11.32 & 1.70 & 0.221 & 0.374 & 7.75 & 40.1 & 4.43 & 14.6  \\
  & 1.3140 & 11.25 & 1.80 & 0.236 & 0.432 & 7.61 & 45.4 & 4.91 & 15.2  \\
\end{tabular}
\end{ruledtabular}
\end{table*}

Finally, in Fig.\ 12 we show both the frequencies and the damping
times of the $w_{I}$-, $w_{II}$-,  $w_{I2}$-, and $w_{I3}$-modes
for canonical neutron stars of mass $M=1.4M_{\odot}$. In order to
evaluate their dependence on the symmetry energy, we used all four
symmetry energy functionals shown in Fig.1, namely $x=0, -0.5, -1$
and the APR, that are within the constraints set by the heavy-ion
reaction data. It is seen that the various modes have different
dependences on the symmetry energy. While the damping time of the
$w_{I}$ mode increases with the increasing frequency as the
symmetry energy becomes softer (from $x=-1$ to $x=0$ as shown in
Fig. 1), the opposite behaviors are observed for the $w_{12}$,
$w_{13}$ and $w_{II}$ modes. Thus, simultaneous studies of
multiple w-modes will be useful in understanding the imprints of
symmetry energy on the w-modes of gravitational waves. It is also
seen that the $w_{II}$-mode has the smallest frequency and damping
time - the frequency of $w_{II}$-mode for a neutron star of mass
$M=1.4M_{\odot}$ with the $x=-1$ EOS is about $1.72
~\textrm{kHz}$.

\section{Summary}
In summary, we have examined the eigen-frequencies of the first few
axial w-modes of oscillating neutron stars by using an EOS with the
symmetry energy partially constrained by the recent terrestrial
nuclear laboratory data. Our studies indicate that the density
dependence of the nuclear symmetry energy $E_{sym}(\rho)$ has a
clear imprint on both the frequency and the damping time of the
axial w-modes. We confirmed the previously found universal behavior
of the mass-scaled eigen-frequencies. Moreover, we explored several
alternative universal scalings of the eigen-frequencies. The latter
scaled with the absolute value of the gravitational energy $|E_{g}|$
are more universal functions independent of the symmetry energy,
especially for the $w_{I2}$-mode. Furthermore, it is found that for
the $w_{II}$-mode to exist neutron stars have to have a minimum
compactness of $M/R\approx 0.1078$ independent of the EOS used.

\begin{acknowledgments}
 The work is supported in part of the National
Natural Science Foundation of China under Grant No. 10647116, the
Young Teachers' Training Program from China Scholarship Council
under Grant No. 2007109651, the US National Science Foundation
under Grants No. PHY0652548 and No. PHY0757839, the Research
Corporation under Award No. 7123 and the Texas Coordinating Board
of Higher Education Grant No.003565-0004-2007.
\end{acknowledgments}

\end{document}